\documentstyle[aps,twocolumn,psfig]{revtex}

\begin{document}
\draft
\title{Some remarks about the Tsallis' Formalism}
\author{L. Velazquez\thanks{%
luisberis@geo.upr.edu.cu}}
\address{Departamento de F\'{i}sica, Universidad de Pinar del R\'{i}o\\
Mart\'{i} 270, esq. 27 de Noviembre, Pinar del R\'{i}o, Cuba. }
\author{F. Guzm\'{a}n\thanks{%
guzman@info.isctn.edu.cu}}
\address{Departamento de F\'{i}sica Nuclear\\
Instituto Superior de Ciencias y Tecnolog\'{i}as Nucleares\\
Quinta de los Molinos. Ave Carlos III y Luaces, Plaza\\
Ciudad de La Habana, Cuba.}
\date{\today}
\maketitle

\begin{abstract}
In the present paper the conditions for the validity of the Tsallis'
Statistics are analyzed. The same has been done following the analogy with
the traditional case: starting from the microcanonical description of the
systems and taking into account their self-similarity scaling properties in
the thermodynamic limit, it is analyzed the necessary conditions for the
equivalence of microcanonical ensemble with the Tsallis' generalization of
the canonical ensemble. It is shown that the Tsallis' Statistics is
appropriate for the macroscopic description of systems with potential
scaling laws of the asymptotic accessible states density of the
microcanonical ensemble. Our analysis shows many details of the Tsallis'
formalism: the q-expectation values, the generalized Legendre's
transformations between the thermodynamic potentials, as well as the
conditions for its validity, having a priori the possibility to estimate the
value of the entropic index without the necessity of appealing to the
computational simulations or the experiment. On the other hand, the
definition of physical temperature received a modification which differs
from the Toral's result. For the case of finite systems, we have generalized
the microcanonical thermostatistics of D. H. E. Gross with the
generalization of the curvature tensor for this kind of description.
\end{abstract}


\section{Introduction}

In the last years many researchers have been working in the justification of
the Tsallis' Formalism. Many of them have pretended to do it in the context
of the Information Theory \cite{Cura,Sumi,Plst} without appealing to the
microscopic properties of the systems. Through the years many functional
forms of the information entropy similar to the Shanonn-Boltzmann-Gibbs'
have been proposed in order to generalize the traditional Thermodynamics
(see for example in refs.\cite{renyi,shamit,abes,lans}). This way to derive
the Thermostatistics is very atractive, since it allows us to obtain
directly the probabilistic distribution function of the generalized
canonical ensemble at the thermodynamic equilibrium, as well as to develop
the dynamical study of systems in non-equilibrium processes.

The main dificulty for this kind of description is to determine the necesary
conditions for the application of each specific entropic form. For example,
in the Tsallis' Statistics, the theory is not be able to determine
univocally the value of the entropy index, $q$, so that, it is needed the
experiment or the computational simulation in order to precise it (see for
example in the refs. \cite{reis,Pla2,tir3}). Similar arguments can be
applied for other formulations of the Thermodynamics based on a parametric
information entropic form. That is the reason why we consider that the
statistical description of nonextensive systems should start from the
microscopic characteristics of them.

Following the traditional analysis, the derivation of the Thermostatistics
from the microscopic properties of the systems could be performed by
considerating the microcanonical ensemble. For the case of the Tsallis'
Statistics this is not a new idea.

In 1994 A. Plastino and A. R. Plastino \cite{Plt2} had proposed one way to
justify the q-generalized canonical ensemble with similar arguments employed
by Gibbs himself in deriving his canonical ensemble. It is based on the
consideration of a closed system composed by a subsystem weakly interacting
with a {\it finite thermal bath}. They showed that the macroscopic
characteristics of the subsystem are described by the Tsallis' potential
distribution, relating the {\it entropy index, }$q${\it ,} with the
finiteness of the last one. In this approach the Tsallis {\it ad-hoc}
cut-off condition comes in a natural fashion.

Another attempt was made by S. Abe and A. K. Rajagopal \cite{Raj1}: a closed
system composed by a subsystem weakly interacting with a very large thermal
bath, this time analyzing the behavior of the systems around the
equilibrium, considering this as a state in which the most probable
configurations are given. They showed that the Tsallis' canonical ensemble
can be obtained if the entropy counting rule is modified, introducing the
Tsallis' generalization of the logarithmic function for arbitrary entropic
index \cite{qln}, showing in this way the possibility of the nonuniqueness
of the canonical ensemble theory.

So far it has been said that the Tsallis' Statistics allows to extend the
Thermodynamics to the study of systems that are anomalous from the
traditional point of view, systems with long-range correlations due to the
presence of long-range interactions, with a dynamics of non markovian
stochastic processes, where the {\it entropic index gives a measure of the
non extensivity degree of a system, an intrinsic characteristic of the same }
\cite{tsal}. The identification of this parameter with the finiteness of a
thermal bath is limited, since this argument is non-applicable on many other
contexts in which the Tsallis' Statistics is expected to work: astrophysical
systems \cite{pla1,pla2}, turbulent fluids and non-screened plasma \cite{bog}%
, etc.

The Abe-Rajogopal's analysis suggests that there is an arbitrariness in the
selection of the\ entropy counting rule, which determines the form of the
distribution. In their works they do not establish a criterium that allows
to define the selection of the entropy counting rule univocally.

In the Boltzmann-Gibbs' Statistics the entropy counting rule is supported by
means of the scaling behavior of the microcanonical states density and the
fundamental macroscopic observables, the integrals of motion and external
parameters, with the increasing of the system degrees of freedom, and its
Thermodynamic Formalism, based on the Legendre's Transformations between the
Thermodynamic potentials,{\em \ by the equivalence between the
microcanonical and the canonical ensembles in the Thermodynamic Limit} (ThL).

In the ref. \cite{vel1} it was addressed the problem of generalizing the
extensive postulates in order to extend the Thermostatistics for some
Hamiltonian non-extensive systems. Our proposition was that this derivation
could be carried out taking into consideration the self-similarity scaling
properties of the systems with the increasing of their degrees of freedom
and analyzing the conditions for the equivalence of the microcanonical
ensemble with the generalized canonical ensemble in the ThL. The last
argument has a most general character than the Gibbs', since it does not
demand the separability of one subsystem from the whole system. The Gibbs'
argument is in disagreement with the long-range correlations of the
nonextensive systems. The consideration of the self-similarity scaling
properties of the systems allows us to precise the counting rule for the
generalized Boltzmann's entropy \cite{vel1}, as well as the equivalence of
the microcanonical ensemble with the generalized canonical one determines
the necesary conditions for the applicability of the generalized canonical
description in the ThL.

\section{The Legendre's Formalism}

In this section the analysis of the necessary conditions for the equivalence
of the microcanonical ensemble with the Tsallis' canonical one will be
performed in analogy with our previous work \cite{vel2}, which was motived
by the methodology used by D. H. E. Gross in deriving his{\em \
Microcanonical Thermostatistics} \cite{gro1} through the technique of the
steepest descend method.

In the ref. \cite{vel2} was shown that the Boltzmann-Gibbs' Statistics can
be applied to the macroscopic study of the pseudoextensive systems, those
with {\em exponential self-similarity scaling laws} \cite{vel1,vel2} in the
ThL, using an adequate selection of the representation of the motion
integrals space \cite{vel1}, $\Im _{N}$. The previous analysis suggests that
a possible application of \ Tsallis' formalism could be found for those
systems with an scaling behavior weaker than the exponential.

In this analysis the following {\it potential self-similarity scaling} laws
will be considered:

\begin{equation}
\left. 
\begin{array}{c}
N\rightarrow N\left( \alpha \right) =\alpha N \\ 
I\rightarrow I\left( \alpha \right) =\alpha ^{\chi }I \\ 
a\rightarrow a\left( \alpha \right) =\alpha ^{\pi _{a}}a
\end{array}
\right\} \Rightarrow W_{asym}\left( \alpha \right) =\alpha ^{\kappa
}W_{asym}\left( 1\right) \text{,}  \label{ps}
\end{equation}
where $W_{asym}$ is the accessible volume of the microcanonical ensemble in
the system configurational space in the ThL, $I$ are the system integrals of
motion of the macroscopic description in a specific representation ${\cal R}%
_{I}$ of $\Im _{N}$, $a$ is a certain set of parameters, $\alpha $ is the
scaling parameter, $\chi $ , $\pi _{a}$ and $\kappa $ are real constants
characterizing the scaling transformations. The nomenclature $W_{asym}\left(
\alpha \right) $ represents:

\begin{equation}
W_{asym}\left( \alpha \right) =W_{asym}\left[ I\left( \alpha \right)
,N\left( \alpha \right) ,a\left( \alpha \right) \right] \text{.}
\end{equation}

This kind of self-similarity scaling laws demands an entropy counting rule
different from the logarithmic. It is supposed that the Tsallis'
generalization of exponential and logarithmic functions \cite{qln}:

\begin{equation}
e_{q}\left( x\right) =\left[ 1+\left( 1-q\right) x\right] ^{\frac{1}{1-q}}%
\text{ \ \ \ }\ln _{q}\left( x\right) =\frac{x^{1-q}-1}{1-q}\equiv
e_{q}^{-1}\left( x\right)
\end{equation}
are more convenient to deal with it.

Let us consider a finite Hamiltonian system with this kind of scaling
behavior in the ThL. We postulate that the {\bf Generalized Boltzmann's
Principle} \cite{vel1} adopts the following form:

\begin{equation}
\left( S_{B}\right) _{q}=\ln _{q}W\text{.}  \label{cr}
\end{equation}
The accessible volume of the microcanonical ensemble in the system
configurational space, $W$, \ is given by:

\begin{equation}
W\left( I,N,a\right) =\Omega \left( I,N,a\right) \delta I_{o}=\delta
I_{o}\int \delta \left[ I-I_{N}\left( X;a\right) \right] dX\text{,}
\end{equation}
where $\delta I_{o}$ is a {\it suitable} constant volume element which makes 
$W$ dimensionless. The corresponding information entropy for the
q-generalized Boltzmann's entropy, the Eq.(\ref{cr}), is the Tsallis'
nonextensive entropy (TNE) \cite{Tsal1}:

\begin{equation}
S_{q}=-%
\mathrel{\mathop{\sum }\limits_{k}}%
p_{k}^{q}\ln _{q}p_{k}\text{.}
\end{equation}

In the thermodynamic equilibrium the TNE leads to the q-exponential
generalization of the Boltzmann-Gibbs' Distributions: 
\begin{equation}
\omega _{q}\left( X;\beta ,N,a\right) =\frac{1}{Z_{q}\left( \beta
,N,a\right) }e_{q}\left[ -\beta \cdot I_{N}\left( X;a\right) \right] \text{,}
\end{equation}
where $Z_{q}\left( \beta ,a,N\right) $ is the partition function \cite{Tsal2}%
 For this ensemble, the {\it q-Generalized Laplace's Transformation} is
given by: 
\begin{equation}
Z_{q}\left( \beta ,N,a\right) =\int e_{q}\left( -\beta \cdot I\right)
W\left( I,N,a\right) \frac{dI}{\delta I_{o}}\text{.}
\end{equation}

The Laplace's Transformation establishes the connection between the
fundamental potentials of both ensembles, the q-generalized {\it Planck's\
potential}:

\begin{equation}
P_{q}\left( \beta ,N,a\right) =-\ln _{q}\left[ Z_{q}\left( \beta ,N,a\right) %
\right] \text{,}
\end{equation}
and the generalized Boltzmann's entropy defined by the Eq.(\ref{cr}):

\begin{equation}
e_{q}\left[ -P_{q}\left( \beta ,N,a\right) \right] =\int e_{q}\left( -\beta
\cdot I\right) e_{q}\left[ \left( S_{B}\right) _{q}\left( I,N,a\right) %
\right] \frac{dI}{\delta I_{o}}\text{.}  \label{e1}
\end{equation}
The q-logarithmic function satisfies the {\it subadditivity relation}:

\begin{equation}
\ln _{q}\left( xy\right) =\ln _{q}\left( x\right) +\ln _{q}\left( y\right)
+\left( 1-q\right) \ln _{q}\left( x\right) \ln _{q}\left( y\right) \text{,}
\end{equation}
and therefore:

\begin{equation}
e_{q}\left( x\right) e_{q}\left( y\right) =e_{q}\left[ x+y+\left( 1-q\right)
xy\right] \text{.}
\end{equation}
The last identity allows us to rewrite the Eq.(\ref{e1}) as:

\begin{equation}
e_{q}\left[ -P_{q}\left( \beta ,N,a\right) \right] =\int e_{q}\left[
-c_{q}\beta \cdot I+\left( S_{B}\right) _{q}\left( I,N,a\right) \right] 
\frac{dI}{\delta I_{o}}\text{,}  \label{e3}
\end{equation}
where:

\begin{equation}
c_{q}=1+\left( 1-q\right) \left( S_{B}\right) _{q}\text{.}
\end{equation}

In the Tsallis' case, the linear form of the Legendre's Transformation {\it %
is violated} and therefore, the{\it \ ordinary Legendre's Formalism does not
establish the correspondence between the two ensembles}. In order to
preserve the homogeneous scaling in the q-exponential function argument, it
must be demanded the scaling invariance of the set of admissible
representations of the integrals of motion space \cite{vel1}, ${\cal M}_{c}$%
, that is, the set ${\cal M}_{c}$ is composed by those representations $%
{\cal R}_{I}$ satisfying the restriction:

\begin{equation}
\chi \equiv 0\text{,}
\end{equation}
in the scaling transformation given in Eq.(\ref{ps}). In these cases, when
the ThL is invoked, the main contribution to the integral of the Eq.(\ref{e3}%
) will come from the maxima of the q-exponential function argument. The
equivalence between the microcanonical and the canonical ensemble will only
take place when there is only one sharp peak. Thus, the argument of the
q-exponential function leads to assume the nonlinear generalization of the
Legendre's Formalism \cite{Abe1,Fran} given by:

\begin{equation}
\widetilde{P}_{q}\left( \beta ,N,a\right) =Max\left[ c_{q}\beta \cdot
I-\left( S_{B}\right) _{q}\left( I,N,a\right) \right] \text{.}  \label{ltp}
\end{equation}

We recognized immediately the formalism of the {\it normalized
q-expectations values }\cite{Abe1}. The maximization leads to the relation:

\begin{equation}
\beta =\frac{\nabla \left( S_{B}\right) _{q}}{1+\left( 1-q\right) \left(
S_{B}\right) _{q}}\left( 1-\left( 1-q\right) \beta \cdot I\right) \text{.}
\label{r1}
\end{equation}
Using the identity:

\begin{equation}
\nabla S_{B}=\frac{\nabla \left( S_{B}\right) _{q}}{1+\left( 1-q\right)
\left( S_{B}\right) _{q}}\text{,}
\end{equation}
where $S_{B}$ is the usual Boltzmann's entropy, the Eq.(\ref{r1}) can be
rewritten as:

\begin{equation}
\beta =\nabla S_{B}\left( 1-\left( 1-q\right) \beta \cdot I\right) .
\end{equation}
Finally it is arrived to the relation:

\begin{equation}
\beta =\frac{\nabla S_{B}}{\left( 1+\left( 1-q\right) I\cdot \nabla
S_{B}\right) }\text{.}  \label{gzl}
\end{equation}

This is a very interesting result because it allows to limit the values of
the entropy index. If this formalism is arbitrarily applied to a
pseudoextensive system (see in ref.\cite{vel2}), then $I\cdot \nabla S_{B}$
will not bound in the ThL and $\beta $ will vanish trivially. The only
possibility in this case is to impose the restriction $q\equiv 1$, that is, 
{\em the Tsallis' Statistics with an arbitrary entropy index can not be
applied to the pseudoextensive systems}. There are many examples in the
literature in which the Tsallis' Statistics has been applied
indiscriminately without minding if the systems are extensive or not, i.e., 
{\it \ gases } \cite{Fran,Abe2}{\it , blackbody radiation }\cite{Tir1}{\it ,
and others}.

In some cases, the authors of these works have introduced some artificial
modifications to the original Tsallis' formalism in order to obtain the same
results as those of the classical Thermodynamics, i.e., the {\it q-dependent
Boltzmann's constant} (see for example in ref.\cite{Abe3}). The above
results indicate the non applicability of the Tsallis' Statistics for these
kind of systems. It must be pointed out that this conclusion is supported
with a great accuracy by direct experimental mensurements trying to find
nonextensive effects in some ordinary extensive systems (cosmic background
blackbody radiation\ \cite{Pla2}, fermion systems \cite{Tsal5,Pla3}, gases 
\cite{Pla4}).

The Tsallis' formalism introduces a correlation to the canonical intensive
parameters of the Boltzmann-Gibbs' Probabilistic Distribution Function. This
result differs from the one obtained by Toral \cite{toral}, who applied to
the microcanonical ensemble the physical definitions of temperature and
pressure introduced by S. Abe in the ref.\cite{Abe2}:

\begin{equation}
\begin{tabular}{l}
$\frac{1}{kT_{phys}}=\frac{1}{1+\left( 1-q\right) S_{q}}\frac{\partial }{%
\partial E}S_{q}$, \\ 
$\frac{P_{phys}}{kT_{phys}}=\frac{1}{1+\left( 1-q\right) S_{q}}\frac{%
\partial }{\partial V}S_{q}$.
\end{tabular}
\end{equation}
When these definitions are applied to the microcanonical ensemble asuming
the generalized Boltzmann's Principle, the Eq.(\ref{cr}), the physical
temperature coincides with the usual Boltzmann's relation:

\[
\frac{1}{kT_{phys}}=\frac{\partial }{\partial E}S_{B}\text{.} 
\]

It is easy to show, that this result does not depend on the entropy counting
rule of the generalized Boltzmann's Principle \cite{vel1}, but on
separability of a closed system in subsystems weakly correlated among them,
and the additivity of the integrals of motion and the macroscopic
parameters. It must be recalled that these exigencies are only valid for the
extensive systems, but, it is not the case that we are studying here. Our
result comes in fashion as consequence of the system scaling laws in the
thermodynamic limit.

An important second condition must be satisfied for the validity of the
Legendre's transformation, {\it the stability of the maximum}. This
condition leads to the q-generalization of the {\em Microcanonical
Thermostatistics} of D. H. E. Gross \cite{gro1}. In this approach, the
stability of the Legendre's formalism is supported by the concavity of the
entropy, the negative definition of the quadratic forms of the curvature
tensor \cite{gro1,vel2}. In the Tsallis' case, the curvature tensor must be
modified as:

\[
\left( K_{q}\right) _{\mu \nu }=\frac{1}{1-\left( 1-q\right) \widetilde{P}%
_{q}}\left[ \left( 2-q\right) \frac{\partial }{\partial I^{\mu }}\frac{%
\partial }{\partial I^{\nu }}\left( S_{B}\right) _{q}+\right. 
\]

\begin{equation}
\left. +\left( 1-q\right) \left( \beta _{\mu }\frac{\partial }{\partial
I^{\nu }}\left( S_{B}\right) _{q}+\beta _{\nu }\frac{\partial }{\partial
I^{\mu }}\left( S_{B}\right) _{q}\right) \right] \text{.}
\end{equation}

Taking into consideration that the scaling behavior of the functions $\left(
S_{B}\right) _{q}$ and $\widetilde{P}_{q}$ are identical, which is derived
from the Eq.(\ref{ltp}), it is easy to see that the curvature tensor is
scaling invariant. Using the above definition and developing the Taylor's
power expansion up to the second order term in the q-exponential argument,
we can approximate the Eq.(\ref{e3}) as:

$e_{q}\left[ -P_{q}\left( \beta ,N,a\right) \right] \simeq \int e_{q}\left[ -%
\widetilde{P}_{q}\left( \beta ,N,a\right) \right] \times $%
\begin{equation}
\times e_{q}\left[ -\frac{1}{2}\left( I-I_{M}\right) ^{\mu }\cdot \left.
\left( -K_{q}\right) _{\mu \nu }\right| _{I=I_{M}}\cdot \left(
I-I_{M}\right) ^{\nu }\right] \frac{dI}{\delta I_{o}}\text{.}  \label{e4}
\end{equation}

The maximum will be stable if all the eingenvalues of the q-curvature tensor
are {\it negative} and {\it very large}. In this case, in the q-generalized
canonical ensemble there will be {\it small fluctuations of the integrals of
motion around its q-expectation values}. The integration of Eq.(\ref{e4})
yields:

$e_{q}\left[ -P_{q}\left( \beta ,N,a\right) \right] \simeq e_{q}\left[ -%
\widetilde{P}_{q}\left( \beta ,N,a\right) \right] \times $%
\begin{equation}
\times \frac{1}{\delta I_{o}\det^{\frac{1}{2}}\left( \frac{1-q}{2\pi }\left.
\left( -K_{q}\right) _{\mu \nu }\right| _{I=I_{M}}\right) }\frac{\Gamma
\left( \frac{2-q}{1-q}\right) }{\Gamma \left( \frac{2-q}{1-q}+\frac{1}{2}%
n\right) }\text{.}  \label{e5}
\end{equation}

Denoting $K_{q}^{-1}$ by:

\begin{equation}
K_{q}^{-1}=\frac{1}{\delta I_{o}\det^{\frac{1}{2}}\left( \frac{1-q}{2\pi }%
\left. \left( -K_{q}\right) _{\mu \nu }\right| _{I=I_{M}}\right) }\frac{%
\Gamma \left( \frac{2-q}{1-q}\right) }{\Gamma \left( \frac{2-q}{1-q}+\frac{1%
}{2}n\right) }\text{,}
\end{equation}
and rewriting Eq.(\ref{e5}) again:

$e_{q}\left[ -P_{q}\left( \beta ,N,a\right) \right] \simeq e_{q}\left[ -%
\widetilde{P}_{q}\left( \beta ,N,a\right) +\right. $%
\begin{equation}
\left. \ln _{q}\left( K_{q}^{-1}\right) -\left( 1-q\right) \ln _{q}\left(
K_{q}^{-1}\right) \widetilde{P}_{q}\left( \beta ,N,a\right) \right] \text{,}
\end{equation}
it is finally arrived to the condition:

\begin{equation}
R\left( q;\beta ,N,a\right) =\left| \left( 1-q\right) \ln _{q}\left(
K_{q}^{-1}\right) \right| \ll 1\text{.}
\end{equation}

The last condition could be considered as an {\it optimization problem},
since the entropic index is an independent variable in the functional
dependency of the physical quantities. The specific value of $q$ could be
chosen in order to minimize the function $R\left( q;\beta ,N,a\right) $ for
all the possible values of the integrals of motion . In this way, the
problem of the determination of the entropic index could be solved in the
frame of the microcanonical theory without appealing to the computational
simulation or the experiment.

Thus, the q-generalized Planck's\ potential could be obtained by means of
the generalized Legendre's transformation:

\begin{equation}
P_{q}\left( \beta ,N,a\right) \simeq c_{q}\beta \cdot I-\left( S_{B}\right)
_{q}\left( I,N,a\right) \text{,}
\end{equation}
where the canonical parameters $\beta $ hold the Eq.(\ref{gzl}). Thus, the
q-generalization of the Boltzmann's entropy will be equivalent with the
Tsallis' entropy in the ThL:

\begin{equation}
\left( S_{B}\right) _{q}\simeq S_{q}\text{.}
\end{equation}

If the uniquenees of the maximum is not guarantized, that is, any of the
eingenvalues of the q-curvature tensor is non negative in a specific region
of the integrals of motion space, there will be a catastrophe in the
generalized Legendre's Transformation. In analogy with the traditional
analysis, this peculiarity can be related with the occurrence of a
phenomenon similar to the phase transition in the ordinary extensive systems.

\section{Conclusions}

We have analyzed the conditions for the validity of the Tsallis'
generalization of the Boltzmann-Gibbs' Statistics. Starting from the
microcanonical ensemble, we have shown that the same one can be valid for
those Hamiltonian systems with potential self-similarity scaling laws in the
asymptotic states density. Systems with this kind of scaling laws must be
composed by strongly correlated particles, and therefore, these systems must
exhibit an anomalous dynamical behavior. There are some computational
evidences that suggest that the Tsallis' Statistics could be applied for
dissipative dynamical systems at the edge of chaos (see in the ref.\cite
{Tsal4,Lato,Lato2,s1,s2,s3,s4,s5}) and Hamiltonian systems with long-range
interactions \cite{hs1,hs2,hs3,hs4}.

In this context we have shown an entire series\ of details of Tsallis'
formalism that in this approach appear in a natural way: the {\it q}%
-expectation values, the generalized Legendre's transformations between the
thermodynamic potentials, as well as the conditions for the validity of the
same one, having a priori the possibility to estimate the value of the
entropic index without the necessity of appealing to the computational
simulations or the experiment.

For the case of finite systems satisfying this kind of scaling laws in the
thermodynamic limit, we have generalized the Microcanonical Thermostatistics
of D. H. E. Gross assuming the Tsallis' generalization of the Boltzmann's
entropy. This assumption leads to the generalization of the curvature
tensor, which is the central object in the thermodynamic formalism of this
theory, since it allows us to access to the ordering information of a finite
system.

\end{document}